\documentclass[reprint, superscriptaddress]{revtex4-2}

\usepackage{amsmath}
\usepackage{amssymb}
\usepackage{amsthm}
\usepackage{amsfonts}
\usepackage{physics}
\usepackage{bbold}

\usepackage{graphicx}
\usepackage[dvipsnames]{xcolor} 
\usepackage{hyperref}
\hypersetup{
    colorlinks = true,
    linkcolor  = red,
    citecolor  = magenta,
    urlcolor   = purple,
    }

\newcommand{\nc}{\newcommand}
\nc{\webirvsp}{\href{https://github.com/zjwang11/irvsp}{\texttt{IRVSP} }}
\nc{\webirtb }{\href{https://github.com/zjwang11/irvsp}{\texttt{IR2TB} }}
\nc{\webUnconMat}{\href{http://tm.iphy.ac.cn/UnconvMat.html}{http://tm.iphy.ac.cn/UnconvMat.html}}
\nc{\webposabr}{\href{https://github.com/zjwang11/UnconvMat/blob/master/src_pos2aBR.tar.gz}{\texttt{POS2ABR} }}

\nc{\ba}{\vb*{a}}
\nc{\bb}{\vb*{b}}
\nc{\bc}{\vb*{c}}
\nc{\bk}{\vb*{k}}
\nc{\bq}{\vb*{q}}
\nc{\hatz}{\hat{z}}
\nc{\bbZ}{\mathbb{Z}}
\nc{\calH}{\mathcal{H}}

\nc{\ie}{{i.e., }} 
\nc{\eg}{{e.g., }} 
\nc{\qhc}{\:\text{H.c.}\:}

\nc{\ii}{\text{i}}
\nc{\eV}{\;\text{eV}}
\nc{\meV}{\;\text{meV}}
\renewcommand{\AA}{\;\text{\AA}}

\nc{\ua}{\uparrow}
\nc{\da}{\downarrow}

\nc{\eq}{=&\;}
\nc{\eqv}{\equiv&\;}
\nc{\dg}[1]{{#1}^{\dagger}}
\nc{\trans}[1]{{#1}^{\scriptscriptstyle{\text{T}}}}

\nc{\q}[1]{Eq. (\ref{#1})}
\nc{\s}[1]{Sec. \ref{#1}}
\nc{\fig}[1]{Fig. \ref{#1}}
\nc{\figref}[2]{Fig. \ref{#1}{\textbf{#2}}}
\nc{\figsref}[2]{Figs. \ref{#1}{\textbf{#2}}}
\nc{\app}[1]{Appendix \ref{#1}}
\nc{\tab}[1]{Table \ref{#1}}
\nc{\Refcite}[1]{Ref. \cite{#1}}

\nc{\green}[1]{{\color{green}{#1}}}
\nc{\blue}[1]{{\color{blue}{#1}}}

\begin{document}
\title{Two elementary band representation model, Fermi surface nesting, and surface topological superconductivity in $A$V$_{3}$Sb$_{5}$ ($A = \text{K, Rb, Cs}$)}

\author{Junze Deng}
\affiliation{Beijing National Laboratory for Condensed Matter Physics,
and Institute of Physics, Chinese Academy of Sciences, Beijing 100190, China}
\affiliation{University of Chinese Academy of Sciences, Beijing 100049, China}

\author{Ruihan Zhang}
\affiliation{Beijing National Laboratory for Condensed Matter Physics,
and Institute of Physics, Chinese Academy of Sciences, Beijing 100190, China}
\affiliation{University of Chinese Academy of Sciences, Beijing 100049, China}

\author{Yue Xie}
\affiliation{Beijing National Laboratory for Condensed Matter Physics,
and Institute of Physics, Chinese Academy of Sciences, Beijing 100190, China}
\affiliation{University of Chinese Academy of Sciences, Beijing 100049, China}

\author{Xianxin Wu}\email{xxwu@itp.ac.cn}
 \affiliation{CAS Key Laboratory of Theoretical Physics, Institute of Theoretical Physics,
Chinese Academy of Sciences, Beijing 100190, China}

\author{Zhijun Wang}\email{wzj@iphy.ac.cn}
\affiliation{Beijing National Laboratory for Condensed Matter Physics,
and Institute of Physics, Chinese Academy of Sciences, Beijing 100190, China}
\affiliation{University of Chinese Academy of Sciences, Beijing 100049, China}

\date{\today}
\begin{abstract}
    The recently discovered vanadium-based Kagome metals $A$V$_{3}$Sb$_{5}$ ($A = \text{K, Rb, Cs}$) are of great interest with the interplay of charge density wave (CDW) order, band topology and superconductivity. 
    In this paper, by identifying elementary band representations (EBRs), we construct a two-EBR graphene-Kagome model to capture the two low-energy van-Hove-singularity dispersions and, more importantly, the nontrivial band topology in these Kagome metals. 
    This model consists of $A_g@3g$ (V-$d_{x^2-y^2/z^2}$, Kagome sites) and  $A_2''@2d$ EBRs (Sb1-$p_z$, honeycomb sites).
    We have investigated the Fermi surface instability by calculating the electronic susceptibility $\chi(\bq)$. 
    Prominent Fermi-surface nesting peaks are obtained at three L points, where the $z$ component of the nesting vector shows intimate relationship with the anticrossing point along M--L. 
    The nesting peaks at L are consistent with the $2\times 2\times 2$ CDW reconstruction in these compounds. 
    In addition, the sublattice-resolved bare susceptibility is calculated and similar sharp peaks are observed at the L points, indicating a strong antiferromagnetic fluctuation.
    Assuming a bulk $s$-wave superconducting pairing, helical surface states and nontrivial superconducting gap are obtained on the (001) surface.
    In analogous to FeTe$_{1-x}$Se$_{x}$ superconductor, our results establish another material realization of a stoichiometric superconductor with nontrivial band topology, providing a promising platform for studying exotic Majorana physics in condensed matter. 
\end{abstract}
\maketitle

\section{Introduction}
Kagome materials have attracted enormous interest in the field of condensed matter due to the possibility of studying the interplay among a frustrated geometry of the crystalline structure \cite{kagome10_Nat, kagome6_PhysRevLett.109.067201, kagome11_Nat, kagome9_science}, nontrivial band topology \cite{kagome1_NatMater, kagome2_PhysRevLett.106.236803, kagome3_PhysRevLett.106.236802, kagome7_sciadv} and unconventional electronic orders \cite{kagome4_PhysRevB.85.144402, kagome5_PhysRevB.87.115135, PhysRevB.86.121105, sc4_PhysRevLett.110.126405}.
Recently, many intriguing phenomena have been reported in vanadium-based Kagome metals $A$V$_{3}$Sb$_{5}$ \cite{exp0_PhysRevMaterials.3.094407, exp1_PhysRevLett.125.247002, exp2_PhysRevMaterials.5.034801, exp3_Yin_2021}, including anomalous Hall effect \cite{ahe1_PhysRevB.104.L041103, ahe2_sciadv.abb6003, ahe3_PhysRevB.105.205104}, pair density wave \cite{pdw1_Nat, pdw2_PhysRevLett.129.167001}, electronic nematicity \cite{nematicity_Nat}, etc.
First-principles calculations and angle-resolved photoemission spectroscopy (ARPES) found the $\bbZ_{2}$-type topology in their band structures \cite{exp1_PhysRevLett.125.247002, HU2022495}.
These Kagome metals host charge density wave (CDW) first-order transition at $T_{\text{CDW}}$ around $80 \sim 104$ K \cite{exp0_PhysRevMaterials.3.094407, exp1_PhysRevLett.125.247002, exp2_PhysRevMaterials.5.034801, exp3_Yin_2021}. 
Various experimental techniques have been applied to determine the $A$V$_3$Sb$_5$ low-temperature structure, including high-resolution X-ray diffraction \cite{cdw_xrd1_PhysRevX.11.031050, cdw_xrd2_PhysRevB.104.195132, cdw_xrd3_PhysRevX.11.041030}, nuclear magnetic resonance and nuclear quadrupole resonance \cite{cdw_nqr1_Mu_2021, cdw_nqr2_scpma}, scanning tunneling microscopy \cite{cdw_stm1_PhysRevX.11.031026, cdw_stm2_NatMater, cdw_stm3_Nat} and ARPES \cite{cdw_arpes_NatCommun,PhysRevB.105.L241111, cdw2_PhysRevResearch.4.033072, PhysRevB.106.L241106} etc.
However, the origin of the CDW phase of $A$V$_3$Sb$_5$ still remains controversial.

Besides the CDW phase transition, $A$V$_3$Sb$_5$ also host superconductivity with a transition temperature $T_{c}$ around $0.9 \sim 2.5$ K at ambient conditions \cite{exp1_PhysRevLett.125.247002, exp2_PhysRevMaterials.5.034801, exp3_Yin_2021}.
To reveal the superconducting gap, multiple experimental techniques have been applied but no consensus has been reached yet. There are accumulating evidences of a fully gapped $s$-wave superconductivity in $A$V$_3$Sb$_5$ \cite{cdw_nqr1_Mu_2021, swave2_Gupta_2022, swave3_scpma}.
Interestingly, the bulk $s$-wave pairing can induce a synthetic $p_x+i p_y$ pairing on helical Dirac surface states and Majorana zero modes (MZMs) may arise inside the vortex cores \cite{tsc1_PhysRevLett.100.096407, tsc2_PhysRevLett.107.097001}.
Surface-dependent zero-bias states with spatial evolution similar to the zero-bias peaks in Bi$_2$Te$_3$/NbSe$_2$ heterostructures \cite{mzm1_PhysRevLett.114.017001} and iron-based superconductors FeTe$_{1-x}$Se$_{x}$ \cite{mzm2_NatPhys, mzm3_PhysRevB.92.115119, mzm4_PhysRevB.93.115129, mzm5_science.aan4596, mzm6_science.aao1797, vortex3_PhysRevLett.117.047001} have been recently resolved in the vortex cores of CsV$_3$Sb$_5$ \cite{cdw_stm1_PhysRevX.11.031026}, suggesting its intrinsic topological superconductivity.
However, the theoretical model description, surface states and spin texture are still lacking for studying surface topological superconductivity in $A$V$_3$Sb$_5$.

In this paper, based on irreducible representations (irreps) and orbital-resolved band structures, we identify that the bands of $A_{g}@3g$ and $A_{2}''@2d$ elementary band representations (EBRs) are crucial for two van Hove singularities (VHSs) near the Fermi level ($E_F$) and the nontrivial band topology.
Although some effective tight-binding (TB) models have been constructed in \Refcite{cdw3_PhysRevLett.127.217601, sc1_PhysRevLett.127.177001, vHS2_NatCommun}, the $p_z$-orbital induced $A_2''@2d$ EBR has long been ignored.
Accordingly, the two-EBR graphene-Kagome model is constructed, which contains $d_{x^2-y^2}$ and $p_z$ orbitals,
to uniquely capture both nontrivial band topology and Fermi surface (FS) nesting.
The FS nesting peaks at three L points are obtained, which are compatible with the $2\times 2\times 2$ reconstruction.
We find that the reconstruction along $k_z$ is intimately related to the band anticrossing point on M--L.
Taking together the bulk $s$-wave superconductivity and surface Dirac-cone states of $A$V$_3$Sb$_5$, our work has established a material realization of a stoichiometric superconductor with topological band structure, providing a promising platform for studying exotic Majorana physics in condensed matter.

\begin{table}[t!]
    \centering
    \caption{The atomic valence-electron band representations of $A$V$_3$Sb$_5$.}
    \label{table:aBRs}
    \begin{ruledtabular}
    \begin{tabular}{ccccrlcc}
        Atom & WKS($q$) & Symm. & Conf. & Irreps($\rho$) & & aBRs($\rho@q$) \\ 
        \hline
        $A$ & $1a$ & 6/mmm & $s^1$ & $s$:&$A_{1g}$ & \\  
        \hline
        V & $3g$ & mmm & $s^2d^3$ & $d_{z^2}$:&$A_g$ & \\
        & & & & $d_{x^2-y^2}$:&$A_g$ & $A_g@3g$ \\
        & & & & $d_{xy}$:&$B_{1g}$ & \\
        & & & & $d_{yz}$:&$B_{3g}$ & $B_{3g}@3g$ \\
        & & & & $d_{zx}$:&$B_{2g}$ & $B_{2g}@3g$ \\ 
        \hline
        Sb1 & $4h$ & 3m & $p^4$ & $p_z$:&$A_1$ & $A_1@4h=$\\
        &&&&&& $(A_{1}'+A_{2}'')@2d$\\
        & & & & $p_x,p_y$:&$E$ &  \\ 
        \hline
        Sb2 & $1b$ & 6/mmm & $p^4$ & $p_z$:&$A_{2u}$ &  $A_{2u}@1b$ \\
        & & & & $p_x,p_y$:&$E_{1u}$ 
        \end{tabular}
    \end{ruledtabular}
\end{table}

\begin{figure}[!b]
    \centering
    \includegraphics[width=0.98\linewidth]{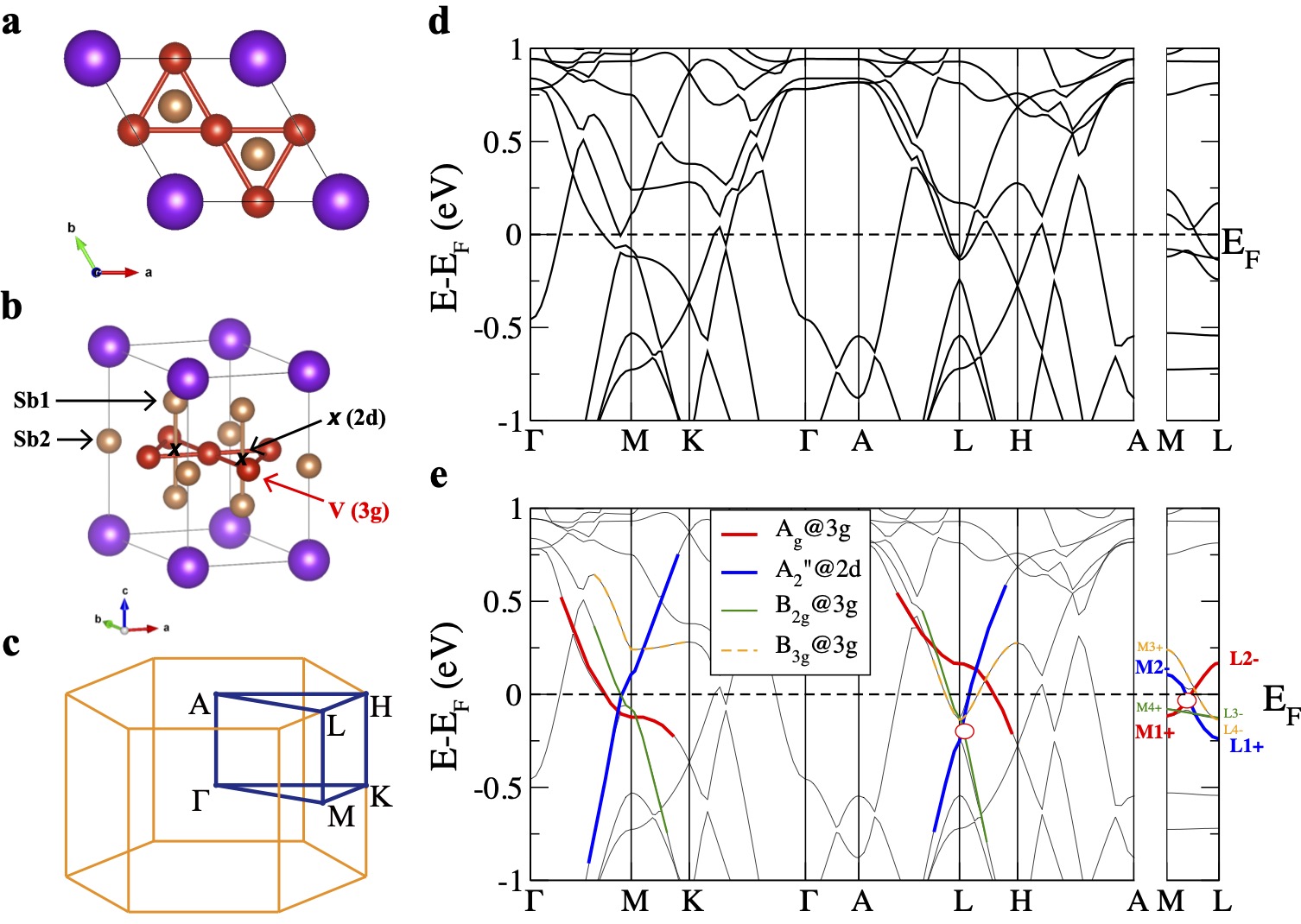}
    \caption{Crystallographic unit cell \textbf{a}, \textbf{b}, Brillouin zone (BZ) \textbf{c} and band structure \textbf{d} of KV$_3$Sb$_5$. 
    According to the irreps, these electronic bands around M/L are classified into four EBRs, highlighted in \textbf{e}.
    The red circles in \textbf{e} denote the hybridization between the highlighted bands.
    }
    \label{fig:dft}
\end{figure}

\begin{figure}[t!]
    \centering
    \includegraphics[width=0.98\linewidth]{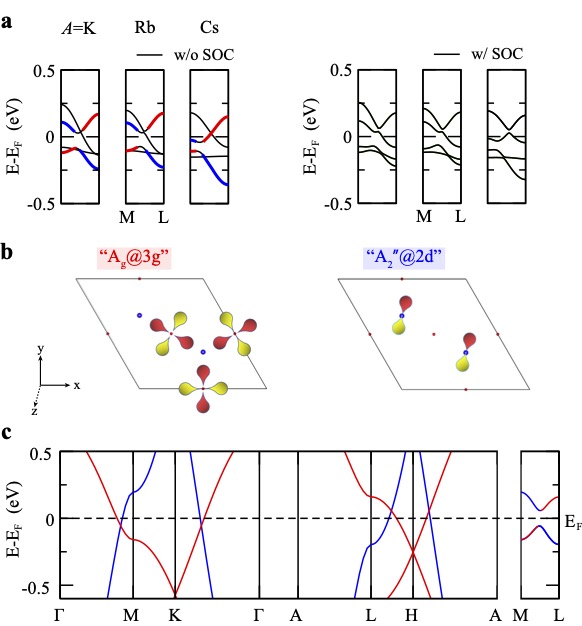}
    \caption{\textbf{a} Band structures of $A$V$_3$Sb$_5$ along M--L without (w/o) SOC and with (w/) SOC.
    \textbf{b} Illustration of the basis orbitals considered in the two-EBR model.
    \textbf{c} The bulk dispersion of the two-EBR model.}
    \label{fig:tb_bands}
\end{figure}

\section{Calculations and results}
\subsection{Crystal structure and band representations}
The pristine phase of $A$V$_{3}$Sb$_{5}$ crystallizes in a layered structure with hexagonal symmetry (SG $P6/mmm$; \#191). 
The $A$, V and Sb1/Sb2 are located at different Wyckoff sites (WKS): $1a$, $3g$ and $4h/1b$. 
Using \webposabr, the atomic valence-electron band representations (aBRs) are presented in \tab{table:aBRs}.
The band structure of the representative KV$_{3}$Sb$_{5}$ is given in \figref{fig:dft}{d}. 
From the irreps and orbital-resolved band structures of Figs. S1\textbf{f-j} in supplemental material (SM) \cite{supmat},
despite the strong hybridization between V-$d_{yz}$ and Sb1-$p_z$ orbitals,
one can still classify low-energy bands at M/L into four EBRs: $A_{g}@3g$ (red; V-$d_{x^2-y^2/z^2}$), $B_{2g}@3g$ (green; V-$d_{xz}$), $B_{3g}@3g$ (orange; V-$d_{yz}$) and $A_{2}''@2d$ (blue; Sb1-$p_z$), as highlighted in \figref{fig:dft}{e}.
The aBR of Sb1-$p_z$ is $A_1@4h$, which is not elementary and can be reduced to two eBRs: $A_1'@2d$ (bonding) and $A_2''@2d$ (anti-bonding).
Based on the obtained irreps, $A_2''@2d$ is crucial to the Fermi-level states.

\begin{figure*}[t!]
    \centering
    \includegraphics[width=0.98\linewidth]{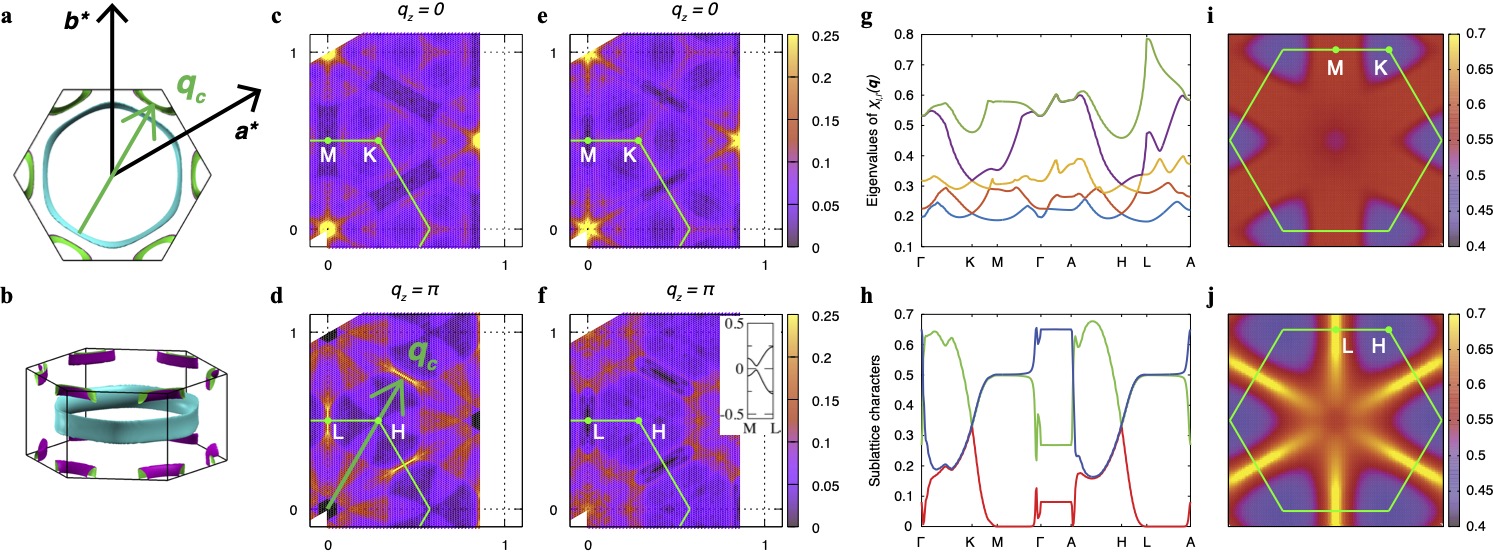}
    \caption{Fermi surfaces \textbf{a}, \textbf{b} and electronic susceptibility \textbf{c}, \textbf{d}.
    The Fermi surface nesting vectors $\bq_c=\frac{1}{2}\ba^*+\frac{1}{2}\bc^*({\rm L}_1),\frac{1}{2}\bb^*+\frac{1}{2}\bc^*({\rm L}_2),\frac{1}{2}\ba^*+\frac{1}{2}\bb^*+\frac{1}{2}\bc^*({\rm L}_3)$ (green arrow) are indicated in panels \textbf{a} and \textbf{d}, where $\ba^*,\bb^*,\bc^*$ are the reciprocal lattice vectors.
    \textbf{e} and \textbf{f} show the computed $\chi(\bq)$ with modified parameters $\varepsilon_d=0.08, \varepsilon_p=0.92, t_3'=-0.008$. The inset of \textbf{f} is the corresponding energy dispersion along M--L.
    \textbf{g} Eigenvalues of the bare susceptibility along high-symmetry $\bq$ paths.
    \textbf{h} Weights of Kagome sublattices for the largest susceptibility eigenvalues and the three colors label the three sublattices.
    Largest susceptibility eigenvalues in the $q_z=0$ plane \textbf{i} and $q_z=\pi$ plane \textbf{j}.
    }
    \label{fig:susceptibility}
\end{figure*}

\subsection{The two-EBR graphene-Kagome model}
A VHS dispersion is formed and becomes relevant due to the presence of a saddle point (at M/L) near the $E_F$.
From \figref{fig:dft}{e}, one can find that the orange $B_{3g}@3g$ band does not form a saddle point at M/L at all. 
Instead, it yields an electron pocket around L. 
Among the other three EBRs, the saddle-point structure of the green $B_{2g}@3g$ EBR is $\sim 200$ meV below the $E_F$ on the whole M--L line (\figref{fig:tb_bands}{a}). 
As a result, it forms a big cylinder hole pocket along $\Gamma$--A. 
In this work, in order to capture the most important band topology, low-energy VHSs and FS nesting, we simply consider a two-EBR model with $A_{g}@3g$ and $A_{2}''@2d$ EBRs (graphene-Kagome model). 
According to aBRs in \tab{table:aBRs}, we conclude that the $A_{g}@3g$ EBR is from V-$d_{x^2-y^2}$ and V-$d_{z^2}$ orbitals, while the $A_{2}''@2d$ EBR is formed by the antibonding state of Sb1-$p_{z}$ orbital.
Interestingly, the $A_g@3g$ BR has been confirmed by earlier ARPES \cite{vHS1_NatPhys, vHS3_NatCommun}, and the Sb1-$p_z$ orbital has been reported responsible for the CDW transition \cite{pz_adma.202209010}.
Although some effective TB models are constructed in the literature \cite{cdw3_PhysRevLett.127.217601, sc1_PhysRevLett.127.177001, vHS2_NatCommun}, the $p_z$ orbital was not included at all. 
Our two-EBR graphene-Kagome model is unique and crucial to capture both FS nesting and the nontrivial band topology.
In addition, a FS of $A_{2u}@1b$ EBR band is from Sb2-$p_{z}$ orbital along $\Gamma$--A (see details in SM \cite{supmat}). 
A complete model with a full set of EBRs can be found in Sec. IV of the SM \cite{supmat}, which reproduces all FSs and respects all the crystal symmetries. It can be used for more detailed calculations and analyses in the future.

With isostructural crystals, the band structures of $A$V$_3$Sb$_5$ are very similar, and the spin-orbit coupling (SOC) effect is weak. 
In the absence of SOC, a hybridization gap already opens between the two EBRs (red and blue bands) on M--L, resulting in an anticrossing point as shown in \figref{fig:tb_bands}{a}. 
The $E_F$ is more or less located within the gap for the three $A$V$_3$Sb$_5$ compounds, while the $k_z$ value of the anti-crossing point varies.
For $A = \text{K, Sb}$ the anticrossing point is near $k_z=\pi/2$, while for $A = \text{Cs}$ it moves to around $k_z=\pi/4$.
The band structure of our graphene-Kagome model is presented in \figref{fig:tb_bands}{c} with the anticrossing point at $k_z=\pi/2$.
In the following, the electronic susceptibility, topological surface states and $s$-wave induced topological superconductivity are studied based on it.

\subsection{Electronic susceptibility and Fermi surface nesting}
To characterize the electronic contribution of the FS from the high-temperature pristine phase of $A$V$_3$Sb$_5$ to the low-temperature CDW phase, the bare electronic susceptibility is calculated. 
In \figsref{fig:susceptibility}{c-f}, we plot the real part of the bare electronic susceptibility in the constant-matrix approximation (Lindhard function),
\begin{equation}\label{eq:sus_charge}
    \begin{aligned}
        \chi(\bq) \eq -\frac{1}{N} \sum_{\bk\mu\nu} \frac{n_{\text{F}}[E_{\mu}(\bk)] - n_{\text{F}}[E_{\nu}(\bk + \bq)]}{E_{\mu}(\bk) - E_{\nu}(\bk + \bq)},
    \end{aligned}
\end{equation}
where $N$ is the number of unit cells, $E_{\nu}(\bk)$ is the eigen energy for the $\nu$th band at the momentum $\bk$ and $n_{\text{F}}(\epsilon) = 1/[\exp(\epsilon/k_{\text B}T)+1]$ denotes the Fermi-Dirac distribution function. 
We observe strong peaks in $\chi(\bq)$ of our model at the three L points in \figref{fig:susceptibility}{d}, which is consistent with the $2\times 2\times 2$ CDW reconstruction \cite{cdw_xrd1_PhysRevX.11.031050}, and is accordant with DFT results \cite{cdw_to_add_PhysRevB.105.155106, cdw_to_add_10.1063/5.0081081}.
Therefore, one can conclude that the two VHSs are crucial to CDW instability.
Once modifying the anticrossing point away from $k_z=\pi/2$ (slightly modifying the parameters), the strong peak at $q_z=\pi$ plane disappears. Instead, the strong peak moves to $\bq_c=(0.5,0.5,0.6)$, which may be related to the observed $2\times 2\times 4$ CDW reconstruction in CsV$_3$Sb$_5$ \cite{cdw_to_add_PhysRevB.105.155106}.
These results reveal that the $q_z$ value of the Fermi-surface nesting vector $\bq_{c}$ is intimately related to the anticrossing point on the M--L path. 

\begin{table}[b!]
    \centering
    \caption{Total energy (meV per formula) of relaxed crystals.}
    \label{table:G}
    \begin{ruledtabular}
    \begin{tabular}{lccc}
        $E_{\text{CDW}} - E_{\text{Pris}}$ & KV$_3$Sb$_5$ & RbV$_3$Sb$_5$ & CsV$_3$Sb$_5$ \\
        \hline
        $2\times 2\times 1$ (SD \cite{cdw1_PhysRevLett.127.046401}) & $-0.15$ & $-2$ & $-4.25$ \\
        $2\times 2\times 1$ (ISD \cite{cdw1_PhysRevLett.127.046401}) & $-1.75$ & $-7.75$ & $-15$ \\
        $2\times 2\times 2$ & $-10.9125$ & $-13.08875$ & $-4.15125$ \\
        $2\times 2\times 4$ &  &  & $-10.159375$
    \end{tabular}
    \end{ruledtabular}
\end{table}

To further check the intrinsic spin fluctuations of the two-EBR TB model, we calculate the sublattice-resolved bare susceptibility with including sublattice weight and the formula is given by,
\begin{equation}\label{eq:sus}
    \begin{aligned}
        \chi^{0}_{l_1 l_2 l_3 l_4}(\bq, \ii\omega_{n}) \eq -\frac{1}{N} \sum_{\bk \mu \nu} a^{l_4}_{\mu}(\bk) a^{l_2 *}_{\mu}(\bk) a^{l_1}_{\nu}(\bk + \bq) \\
        & \times a^{l_3 *}_{\nu}(\bk + \bq) \frac{n_{\text{F}}[E_{\mu}(\bk)] - n_{\text{F}}[E_{\nu}(\bk + \bq)]}{\ii\omega_{n} + E_{\mu}(\bk) - E_{\nu}(\bk + \bq)},
    \end{aligned}
\end{equation}
where $a^{l_i}_{\mu}(\bk)$ is the $l_i$th component of the eigenvector for band $\mu$ resulting from the diagonalization of the TB Hamiltonian.
The eigenvalues of the static bare susceptibility matrix $\chi_{l,l'}(\bq)\equiv\chi^{0}_{lll'l'}(\bq, 0)$ along high-symmetry paths are displayed in \figref{fig:susceptibility}{g}. 
A sharp peak develops at the L point for the largest eigenvalue, indicating a strong antiferromagnetic (AFM) fluctuation. 
According to the eigenvectors, we find that the largest eigenvalue is dominantly contributed by the $A_g@3g$ EBR (the $A_g$-irrep orbitals on the Kagome lattice), due to the relative flatness (or narrow band width) of its VHS structure. 
We further plot the corresponding weights of three sublattices in \figref{fig:susceptibility}{h}, where the peak of the bare susceptibility at L is primarily attributed to two sublattices of the Kagome sites, reflecting the pure-sublattice ($p$-type) nature of the VHS \cite{sc1_PhysRevLett.127.177001}. 
\figsref{fig:susceptibility}{i} and \textbf{j} show the largest eigenvalues in $q_z=0$ plane and $q_z=\pi$ plane, respectively. The high intensity around the L point is similar to the Lindhard function, indicating prominent intraband nesting in our two-EBR model.

\subsection{$2\times 2\times 2$ and $2\times 2\times 4$ CDW reconstructions}
The nesting peaks in the susceptibility are correlated with the band anticrossing in the M--L line, originating from the $A_2''@2d$-EBR band shifting in three compounds (\figref{fig:tb_bands}{a}). Motivated by the experimental detection of different CDW configurations $2\times2\times2$ or $2\times2\times4$ \cite{cdw_stm1_PhysRevX.11.031026, cdw_xrd3_PhysRevX.11.041030, cdw_PhysRevB.105.195136}, we study the effect of distortions of Sb1 atoms along the $z$ direction on the CDW configurations. 
To initialize the $2\times 2\times 2$ and $2\times 2\times 4$ supercell structures, certain Sb1 atoms are shifted up/down slightly from the pristine structure as indicated in Figs. S4\textbf{b} and \textbf{c} in SM \cite{supmat}.
After the full relaxation in our first-principles calculations, the $2\times 2\times 2$ supercell yields an intertwined structure with alternating SD and ISD distortions, as shown in Fig. S4 \textbf{d} in SM \cite{supmat}.
However, the $2\times 2\times 4$ supercell yields a super structure with 1/4 ISD and 3/4 SD distortions for CsV$_3$Sb$_5$ (Fig. S4\textbf{e} in SM \cite{supmat}), while it returns to the $2\times 2\times 2$ intertwined structure for K, Rb.
The energy saving for different CDW distortions relative to the pristine structure are tabulated in \tab{table:G}. 
They suggest that the $2\times 2\times 2$ reconstruction is more favourable for $A = \text{K, Rb}$ while the $2\times 2\times 4$ reconstruction is more favourable for $A = \text{Cs}$, consistent with experimental observations \cite{cdw_stm1_PhysRevX.11.031026, cdw_xrd3_PhysRevX.11.041030, cdw_PhysRevB.105.195136}.

\subsection{Topological surface states and spin texture}
\begin{figure}[t!]
    \centering
    \includegraphics[width=0.98\linewidth]{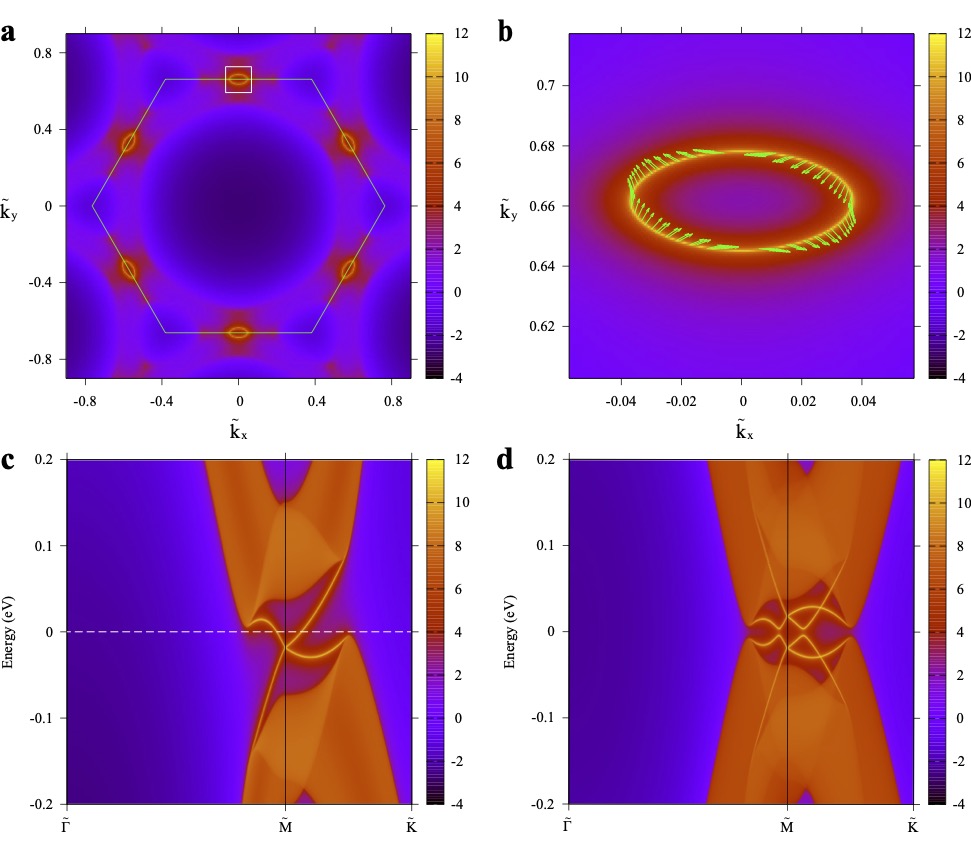}
    \caption{The Fermi surface contour \textbf{a} and the spin texture \textbf{b} of zoom-in area of the (001) surface. 
    There are three FS ovals around $\widetilde{\rm M}$.
    The normal band energy spectrum \textbf{c}, and the superconducting energy spectrum \textbf{d} of the (001) surface with an $s$-wave superconducting gap.}
    \label{fig:surf_states}
\end{figure}

The two EBRs belong to different $M_z$ (reflection with respect to the $xy$ plane) eigenvalues, resulting in a nodal-line semimetal in the absence of SOC.
Upon including SOC ($t_{so}$ in Table S1), it is fully gapped and belongs to a topological insulator (TI) phase, with $\bbZ_2=1$ by parity criterion. 
A surface Dirac cone is obtained at $\widetilde{\text{M}}$ in the (001) surface (\figref{fig:surf_states}{a}), in agreement with the ARPES and DFT results \cite{exp1_PhysRevLett.125.247002}.
A spin-momentum-locked helical spin texture is found for the topological surface states, as shown by the green arrows in the zoom-in plot of \figref{fig:surf_states}{b}.
The $C_{2z}T$ symmetry in the (001) surface restricts $\ev*{\hat{S}_{z}}$ to be zero at arbitrary $(\tilde{k}_{x}, \tilde{k}_{y})$ in the reduced BZ.
Different from that of the $p-p$ band inversion in strong TI Bi$_2$Se$_3$ \cite{spintext1_NatPhys, spintext2_PhysRevLett.105.266806, spintext3_science.1167733, spintext4_PhysRevLett.106.216803, spintext5_PhysRevLett.106.257004, spintext6_PhysRevB.84.165113}, the spin texture originates from the $d-p$ band inversion in $A$V$_3$Sb$_5$ \cite{spintext_PhysRevB.93.205303}.

\subsection{Surface topological superconductivity}
Finally we discuss the potential surface topological superconductivity of $A$V$_3$Sb$_5$. 
We have demonstrated that $A$V$_3$Sb$_5$ possesses a helical surface Dirac cone at $\widetilde{\text{M}}$. 
Since $A$V$_3$Sb$_5$ is intrinsically superconducting without the aid of doping or artificial heterostructures \cite{exp0_PhysRevMaterials.3.094407, exp1_PhysRevLett.125.247002, exp2_PhysRevMaterials.5.034801, exp3_Yin_2021}, 
the topological surface states can be superconducting below $T_c$ as a consequence of surface-bulk proximity effect. 
Then, we introduce an on-site intraorbital spin-singlet bulk pairing and perform a simulation with surface Green's function method for a semi-infinite (001) slab to obtain the surface states (\figref{fig:surf_states}{c}).
It is apparent from \figref{fig:surf_states}{d} that, apart from the bulk superconducting gap, the surface states also obtain an isotropic gap with a gap size, close to the bulk value. 
It is topological with an effective $p_x+ i p_y$ paring, since there are three FS ovals around three $\widetilde{\rm M}$ \cite{xy_PhysRevB.81.134508, xy_PhysRevB.81.220504, xy_PhysRevLett.105.097001}. 
This surface topological superconductivity can be destroyed by varying the chemical potential, which is sensitive to the surface terminations \cite{cdw_stm1_PhysRevX.11.031026}. 
Therefore, $A$V$_3$Sb$_5$, an intrinsic superconductor with topological Dirac-cone surface states, is a potential platform to study MZMs in the core of vortices \cite{tsc1_PhysRevLett.100.096407, tsc2_PhysRevLett.107.097001}.


\section{Conclusion}
In this work, we identify that the $A_{g}@3g$ and $A_{2}''@2d$ VHSs are important to the FS nesting and nontrivial band topology.
The two-EBR graphene-Kagome model is constructed from $d_{x^2-y^2}$ and long-neglected $p_{z}$ orbitals.
The FS nesting peaks at three L points are obtained, which are compatible with the $2\times 2\times 2$ reconstruction. 
The CDW reconstruction along the $z$ direction is intimately correlated with the band anticrossing point on the M--L line, causing a favored $2\times 2\times 2$ reconstruction for $A = \text{K, Rb}$ while the $2\times 2\times 4$ reconstruction is more favourable for $A = \text{Cs}$.
A strong AFM fluctuation is implied from the sublattice-resolved bare susceptibility calculations.
The largest eigenvalue of bare susceptibility matrix $\chi_{l,l'}(\bq)$ is dominantly contributed by the $A_g$ orbitals on the Kagome lattice, due to the relative flatness of its VHS structure.
The nontrivial topology of $A$V$_3$Sb$_5$ can be well reproduced by the graphene-Kagome model, with a surface Dirac cone at $\widetilde{\text{M}}$.
By combining nontrivial band topology and superconductivity, this type of intrinsic topological superconductors holds great promise for studying fascinating aspects of topological superconductors such as MZMs.

\begin{acknowledgments}
This work was supported by the National Key R\&D Program of China (Grant No. 2022YFA1403800), the National Natural Science Foundation of China (Grants No. 11974395, No. 12188101 and No. 12047503), the Strategic Priority Research Program of Chinese Academy of Sciences (Grant No. XDB33000000), and the Center for Materials Genome.
\end{acknowledgments}

\bibliographystyle{apsrev4-1}

%

\clearpage
\begin{widetext}

\setcounter{section}{0}
\setcounter{subsection}{0}
\setcounter{equation}{0}
\setcounter{figure}{0}
\setcounter{table}{0}

\renewcommand{\thesubsection}{\Alph{subsection}}
\renewcommand{\theequation}{S\arabic{equation}}
\renewcommand{\thetable}{S\arabic{table}}
\renewcommand{\thefigure}{S\arabic{figure}}

{\large \textbf{Supplemental material for ``Two elementary band representation model, Fermi surface nesting, and surface topological superconductivity in $A$V$_{3}$Sb$_{5}$ ($A = \text{K, Rb, Cs}$)"}}

\section{\label{app:dft}Calculation methods}
We performed the first-principles calculations based on the density functional theory (DFT) using projector augmented wave (PAW) method \cite{PAW1, PAW2} implemented in the Vienna \emph{ab initio} simulation package (VASP) \cite{VASP1, VASP2} to obtain the electronic structures.
The generalized gradient approximation (GGA) with exchange-correlation functional of Perdew, Burke and Ernzerhof (PBE) for the exchange-correlation functional \cite{PBE} was adopted.
The kinetic energy cutoff  was set to $500$ eV for the plane wave bases.
The BZ was sampled by $\Gamma$-centered Monkhorst-Pack method \cite{MPmethod} with a $12 \times 12 \times 6$ $\bk$-mesh in the self-consistent process.
The irreducible representations of electronic states are obtained by IRVSP \cite{IRVSP}.
The BR analysis is done on the UnconvMat website \cite{webUnconvMat}.

\section{\label{app:local_orbs}Local orbitals and orbital-resolved band structures}
The V-$d$ orbitals in $A$V$_{3}$Sb$_{5}$ split into five irreps under the site-symmetry group (due to the crystal field; Table I in the main text), of which the corresponding basis functions are defined in \figsref{fig:local_orbs}{a-c}.
The orbital-resolved band structures of these local orbitals are presented in \figsref{fig:local_orbs}{f-h}.
The orbital-resolved band structures of the $p_{z}$ orbitals of Sb1 ($A_2''@2d$) and Sb2 ($A_{2u}@1b$) are presented in \figsref{fig:local_orbs}{i} and \textbf{j}.


\begin{figure}[ht]
    \centering
    \includegraphics[width=0.98\linewidth]{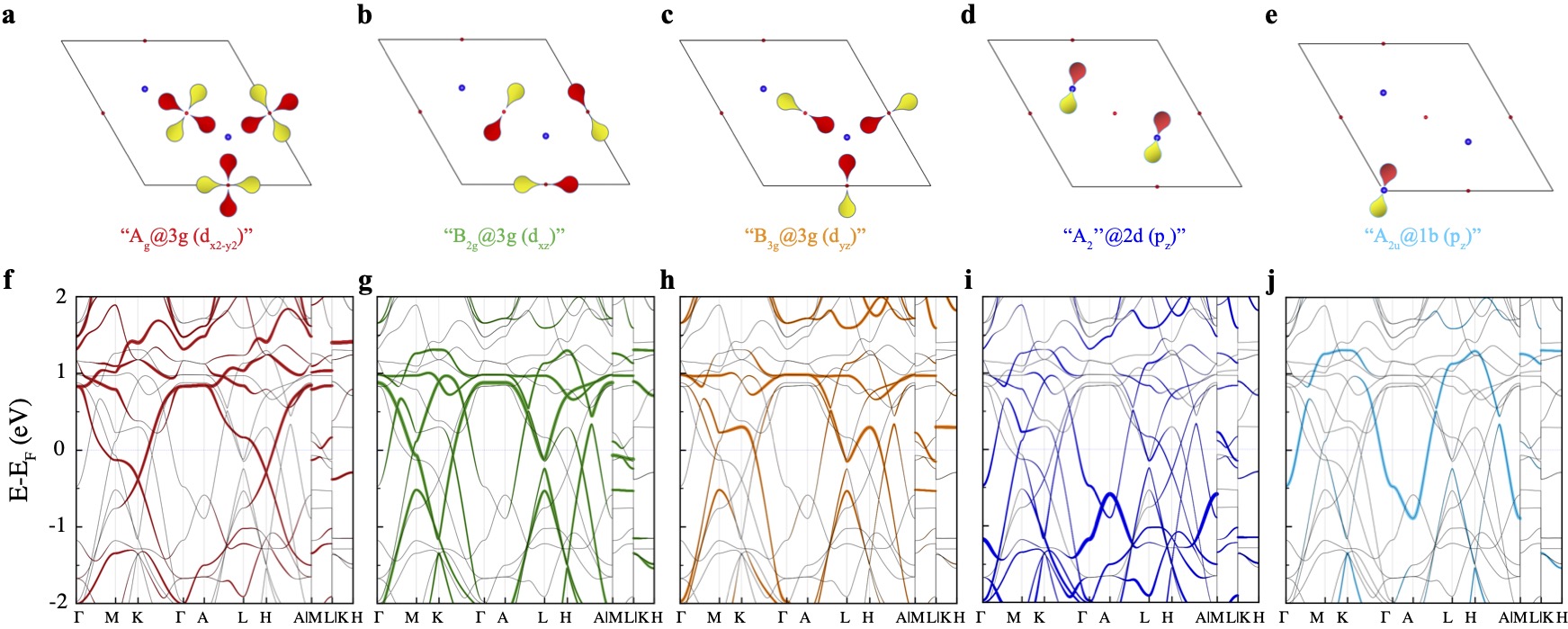}
    \caption{
    Illustrations of local orbitals defined as the basis of corresponding EBRs in KV$_{3}$Sb$_{5}$, \textbf{a} $A_{g}@3g$, \textbf{b} $B_{2g}@3g$, \textbf{c} $B_{3g}@3g$, \textbf{d} $A_{2}''@2d$ and \textbf{e} $A_{2u}@1b$, and their orbital-resolved band structures \textbf{f}, \textbf{g}, \textbf{h}, \textbf{i} and \textbf{j}.}
    \label{fig:local_orbs}
\end{figure}

\section{\label{app:tb_models}Details of the two-EBR TB model}
We construct the two-EBR TB model with considering \emph{local} V-$d_{x^{2}-y^{2}}$ orbitals locating at $3g$ WKS and Sb-$p_{z}$ locating at $2d$ WKS.
The hopping terms considered are illustrated in \fig{fig:tb_hoppings}, and the corresponding parameters are listed in \tab{tab:hop_parameters}.
All hopping terms included in the two-EBR model can be derived under symmetry constraints of SG P6/mmm (\# 191).
The corresponding Hamiltonian then explicitly reads,
\begin{equation}
    \begin{aligned}
        H \eq \sum_{\bk} \calH(\bk) \dg{c}_{\bk} c_{\bk},
        \quad \calH(\bk) = \mqty(h(\bk) & -\lambda(-\bk)^{*} \\ \lambda(\bk) & h(-\bk)^{*})
    \end{aligned}
\end{equation}
where
\begin{equation}
    \begin{aligned}
        h(\bk) \eq 
        \begin{pmatrix}
            \varepsilon_{d} + 2 t_{2}' \cos(\bk\cdot\ba_{3}) & t_{2} ( 1 + e^{\ii\bk\cdot\ba_{2}} ) & t_{2} ( 1 + e^{\ii\bk\cdot\ba_{1}} ) & 2\ii t_{3}' \sin(\bk\cdot\ba_{3}) & 2\ii t_{3}' \sin(\bk\cdot\ba_{3}) \\
            & \varepsilon_{d} + 2 t_{2}' \cos(\bk\cdot\ba_{3}) & t_{2} ( e^{\ii\bk\cdot\ba_{1}} + e^{-\ii\bk\cdot\ba_{2}} ) & 2\ii t_{3}' \sin(\bk\cdot\ba_{3}) e^{-\ii\bk\cdot\ba_{2}} & 2\ii t_{3}' \sin(\bk\cdot\ba_{3}) \\
            & & \varepsilon_{d} + 2 t_{2}' \cos(\bk\cdot\ba_{3}) & 2\ii t_{3}' \sin(\bk\cdot\ba_{3}) & 2\ii t_{3}' \sin(\bk\cdot\ba_{3}) e^{-\ii\bk\cdot\ba_{1}} \\
            & \dagger & & \varepsilon_{p} + 2 t_{1}' \cos(\bk\cdot\ba_{3}) & t_{1} ( 1 + e^{\ii\bk\cdot\ba_{2}} + e^{-\ii\bk\cdot\ba_{1}} ) \\
            & & & & \varepsilon_{p} + 2 t_{1}' \cos(\bk\cdot\ba_{3})
        \end{pmatrix}
    \end{aligned}
\end{equation}
and
\begin{equation}
    \begin{aligned}
        \lambda(\bk) \eq t_{so}
        \begin{pmatrix}
            & & & e^{\ii\frac{-\pi}{6}} & e^{\ii\frac{5\pi}{6}} \\
            &\mathbb{0}_{3\times 3} & & e^{\ii\frac{\pi}{2}} e^{-\ii\bk\cdot\ba_{2}} & e^{\ii\frac{-\pi}{2}} \\
            & & & e^{\ii\frac{-5\pi}{6}} & e^{\ii\frac{\pi}{6}} e^{-\ii\bk\cdot\ba_{1}} \\
            e^{\ii\frac{5\pi}{6}} & e^{\ii\frac{-\pi}{2}} e^{\ii\bk\cdot\ba_{2}} & e^{\ii\frac{\pi}{6}}  \\
            e^{\ii\frac{-\pi}{6}} & e^{\ii\frac{\pi}{2}} & e^{\ii\frac{-5\pi}{6}} e^{\ii\bk\cdot\ba_{1}} &&\mathbb{0}_{2\times 2}
        \end{pmatrix}
    \end{aligned}
\end{equation}
where $\ba_{i}$ is the direct lattice vectors,
\begin{equation}
    \begin{aligned}
        \mqty(\ba_{1} & \ba_{2} & \ba_{3}) \eq \mqty(a & -\frac{a}{2} & 0 \\ 0 & \frac{\sqrt{3}a}{2} & 0 \\ 0 & 0 & c)
    \end{aligned}
\end{equation}


\begin{figure}[ht]
    \centering
    \includegraphics[width=0.4\linewidth]{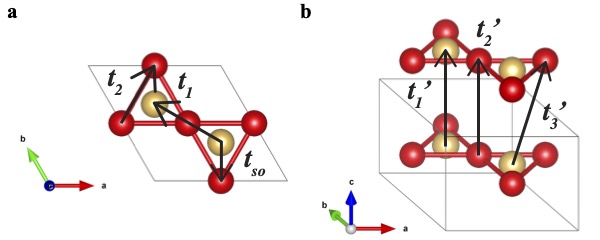}
    \caption{Hopping terms considered in the two-EBR model.}
    \label{fig:tb_hoppings}
\end{figure}

\begin{table}[b]
    \centering
    \caption{The fitting parameters (in eV) of the two-EBR model for $A$V$_{3}$Sb$_{5}$.}
    \label{tab:hop_parameters}
    \begin{ruledtabular}
    \begin{tabular}{c c c c c c c c }
        $\varepsilon_{d}$ & $\varepsilon_{p}$ & $t_{1}$ & $t_{2}$ & $t_{1}'$ & $t_{2}'$ & $t_{3}'$ & $t_{so}$ \\
        \hline
        $0$ & $1$ & $1$ & $-0.414$ & $0.098$ & $-0.08$ & $-0.02$ & $0.09$
    \end{tabular}
    \end{ruledtabular}
\end{table}

\section{\label{app:dft_fit}The complete model with a full set of EBRs for $\text{KV$_3$Sb$_5$}$}
The band representations of the bands around Fermi level are solved via POS2ABR \cite{webposabr} as illustrated in \app{app:local_orbs}.
We then construct tight-binding models according to these EBRs with hopping terms shown in \figsref{fig:hyb_1}{a-d}, of which the fitting parameters are listed in \tab{tab:dft_fit}.
\figsref{fig:hyb_1}{e} and \textbf{f} are the combined band structures of the tight-binding models consist of $A_g@3g$, $B_{2g}@3g$, $A_2''@2d$, and $A_{2u}@1b$, excluding and including the hybridization. The irreps of the tight-binding bands at Fermi level are identical to that from DFT calculations \cite{IRVSP}.

\begin{table}[!ht]
    \centering
    \caption{The fitting parameters (in eV) for the tight-binding band structures to KV$_3$Sb$_5$.}
    \label{tab:dft_fit}
    \begin{tabular}{c|c|c|c|c}
    \hline\hline
    \multicolumn{5}{c}{intra-EBR hopping terms} \\
    \hline
    \figref{fig:hyb_1}{e} & $A_{g}@3g \; (x=g)$ & $B_{2g}@3g \; (x=t)$ & $A_{2}''@2d \; (x=d)$ & $A_{2u}@1b \; (x=b)$ \\
    \hline
    $x_0$   & $ 0.045$ & $-0.7852$ & $0.9085$ & $1.15$   \\
    $x_0^z$ & $-0.072$ & $ 0.0473$ & $0.0878$ & $ 0.015$ \\
    $x_1$   & $-0.41$  & $ 0.352$  & $0.9555$ & $-0.3$   \\
    $x_1^z$ & $ 0.015$ & ---       & ---      & $ 0.015$ \\
    $x_2$   & $-0.034$ & $ 0.1$    & ---      & ---      \\
    $x_2^z$ & $-0.017$ & ---       & ---      & ---      \\
    $x_3$   & ---      & $ 0.352$  & ---      & ---      \\
    $x_3^z$ & ---      & $-0.0176$ & ---      & ---      \\
    \hline\hline
    \multicolumn{5}{c}{inter-EBR hopping terms (hybridization)} \\
    \hline
    \figref{fig:hyb_1}{f} & \multicolumn{4}{c}{$\Delta_{B_{2g} , A_{2}''} = 0.25, \;\; \Delta_{A_{g} , A_{2}''}= -0.02, \;\; d_0 \to 0.4005$} \\
    \hline\hline
    \end{tabular}
\end{table}

\begin{figure*}
    \centering
    \includegraphics[width=0.95\linewidth]{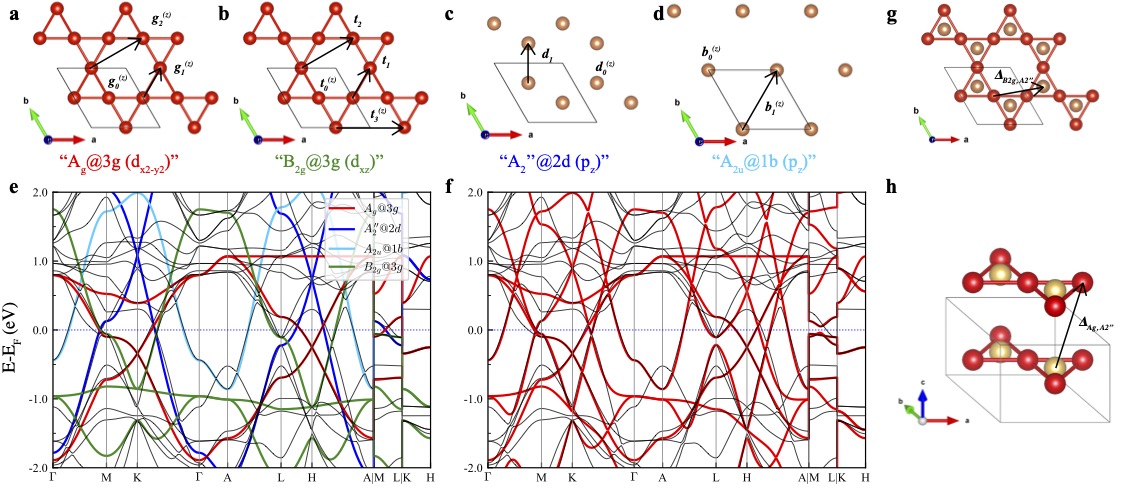}
    \caption{
    Illustration of intra-EBR hopping terms considered in the tight-binding model \textbf{a} $A_{g}@3g$, \textbf{b} $B_{2g}@3g$, \textbf{c} $A_{2}''@2d$ and \textbf{d} $A_{2u}@1b$.
    Band structures from DFT (black solid lines) v.s. a tight-binding model containing EBRs $A_g@3g$, $B_{2g}@3g$, $A_2''@2d$, and $A_{2u}@1b$, without \textbf{e} and with \textbf{f} hybridization.
    Illustration of hybridization terms (inter-EBR hopping terms) considered in panel \textbf{f}, $\Delta_{B_{2g}, A_{2}''}$ \textbf{g} and $\Delta_{A_g, A_{2}''}$ \textbf{h}.}
    \label{fig:hyb_1}
\end{figure*}

\section{\label{app:cdw_struc}The CDW reconstructed lattice structure of $\text{$A$V$_3$Sb$_5$}$}

Since the $A_2''@2d$ EBR is formed by the Sb1 atoms, we consider to modify Sb1 atoms to reconstruct the CDW supercells. 
To initialize the $2\times 2\times 2$ and $2\times 2\times 4$ structures, certain Sb1 atoms are shifted up/down slightly from the pristine structure as indicated in \figsref{fig:cdw_structs}{b} and \textbf{c}.
After full relaxation in DFT calculations, the $2\times 2\times 2$ supercell yields an intertwined structure with alternating SD and ISD, as shown in \figref{fig:cdw_structs}{d}. 
However, the $2\times 2\times 4$ supercell yields a super structure with 1/4 ISD and 3/4 SD for CsV$_3$Sb$_5$ (\figref{fig:cdw_structs}{e}), while it becomes the $2\times 2\times 2$ intertwined structure for K, Rb.
The energy conservations are tabulated in Table II in the main text. 
They suggest that the $2\times 2\times 2$ reconstruction is more favourable for $A = \text{K, Rb}$ while the $2\times 2\times 4$ reconstruction is more favourable for $A = \text{Cs}$.

\begin{figure*}
    \centering
    \includegraphics[width=0.98\linewidth]{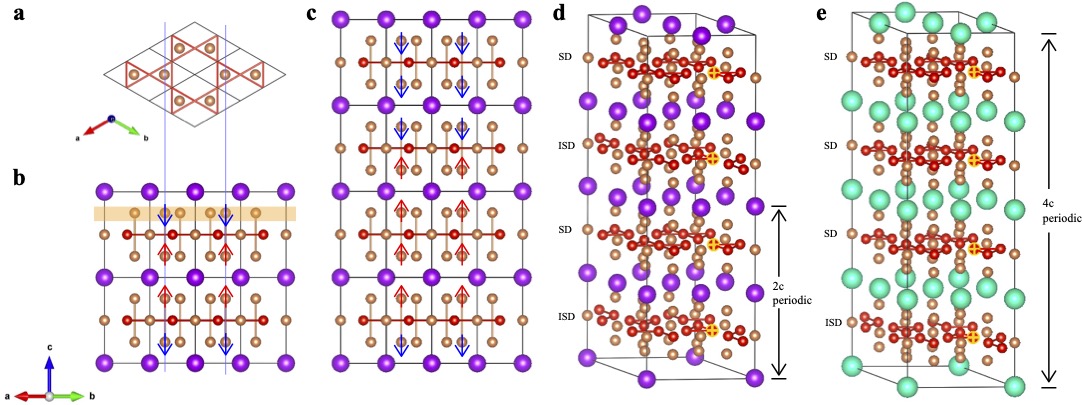}
    \caption{The modified structures \textbf{a}, \textbf{b}, \textbf{c} and relaxed structures \textbf{d}, \textbf{e} of $2\times 2\times 2$ and $2\times 2\times 4$ reconstructions. 
    The modified Sb1 layer is clearly shown in panel \textbf{a}. The displacement in $z$ direction is indicated by the arrows. After full relaxation of the $2\times 2\times 4$ reconstruction, it is $4c$ periodic for CsV$_3$Sb$_5$, while it becomes $2c$ periodic for KV$_3$Sb$_5$ and RbV$_3$Sb$_5$.}
    \label{fig:cdw_structs}
\end{figure*}


\end{widetext}

\end{document}